\documentclass{moriond}
\usepackage{amsmath}

\def\JournalRef#1#2#3#4#5{\href{#5}{{#1} {\bf #2}, #3 (#4)}}

\def\NIMA{{\em Nucl.~Instrum.~Methods}~A}

\def\PRL{\em Phys.~Rev.~Lett.}
\def\PRD{{\em Phys.~Rev.}~D}

\def\EPJC{{\em Eur.~Phys.~J.}~C}
\def\PTEP{\em Prog.~Theor.~Exp.~Phys.}
\def\JHEP{\em J.~High~Energ.~Phys.}
\def\ARNPS{\em Ann.~Rev. Nucl.~Part.~Sci.}
\def\JPG{{\em J.~Phys.}~G}

\def\PR{\em Phys.~Rept.}

\def\be{\begin{equation}}
\def\ee{\end{equation}}
\def\bea{\begin{eqnarray}}
\def\eea{\end{eqnarray}}

\def\un#1{\,\mathrm{#1}}
\def\GeV{\un{GeV}}
\def\MeV{\un{MeV}}

\def\deg{{}^\circ}
\def\rad{\un{rad}}
\def\cm{\un{cm}}
\def\pip{\pi^+}
\def\pim{\pi^-}
\def\kp{K^+}
\def\km{K^-}

\def\Dchi{\Delta \chi^2}
\def\Dm{\Delta m^2}
\def\Dmtt{\Dm_{32}}
\def\Dmbar{\Delta \overline{m}^2}
\def\Dmttbar{\Dmbar_{32}}
\def\dcp{\delta_\mathrm{CP}}
\def\sinot{\sin^2\theta_{13}}
\def\sintt{\sin^2\theta_{23}}
\def\sinttt{\sin^22\theta_{23}}
\def\sinttbar{\sin^2\overline{\theta}_{23}}
\def\tot{\theta_{13}}
\def\ttt{\theta_{23}}
\def\nubar{\overline{\nu}}
\def\numu{\nu_\mu}
\def\numubar{\nubar_\mu}
\def\nue{\nu_e}
\def\nuebar{\nubar_e}

\newcommand{\Photo}{\includegraphics[height=35mm]{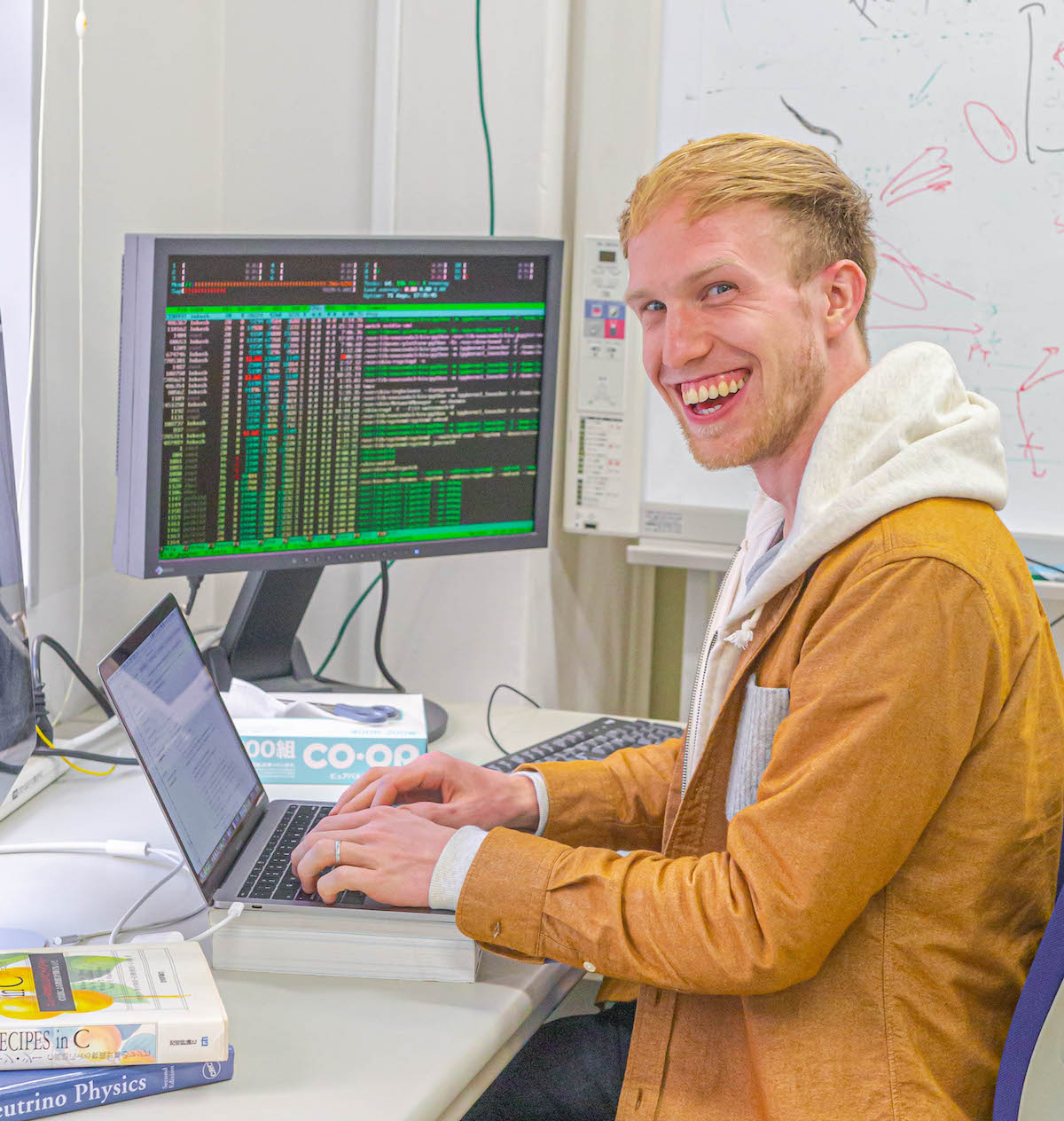}}

\begin{document}
\vspace*{4cm}
\title{RECENT RESULTS FROM T2K}

\author{ L. BERNS for the T2K collaboration }

\address{Department of Physics, Tokyo Institute of Technology,\\2--12--1 Ookayama, Meguro-ku, 152--8551 Tokyo, Japan}

\maketitle\abstracts{
We report on latest results from the Tokai-to-Kamioka experiment studying the disappearance of $\numu$ and appearance of $\nue$ from neutrino oscillation with accelerator generated neutrinos.
Thanks to increase in statistics and numerous upgrades to each part of the oscillation analysis, we set the most sensitive constraints on some of the oscillation parameters, including in particular the CP-violation angle $\dcp$. Future prospects are also briefly discussed.
}

\section{Introduction}

The Tokai-to-Kamioka (T2K) experiment\,\cite{t2k-experiment} is a long baseline neutrino experiment in Japan. A powerful beam of neutrinos produced at the J-PARC accelerator in Tokai is sent to the Super-Kamiokande (SuperK) detector in Kamioka $295~\mathrm{km}$ away. Because for neutrinos the interaction basis is different from the Hamiltonian basis, the neutrino flavor oscillates over propagation distance $L$ with frequency controlled by squared mass splittings over energy $\Dm/E$, and amplitude by the mixing matrix $U$ parametrized by $\theta_{12},\theta_{13},\theta_{23},$ and $\dcp$. One of the interesting open questions in high energy physics is the possibility of CP violation in leptonic sector via non-zero value of $\sin\dcp$. In T2K, a beam composed mostly of $\numu$ (neutrino mode) or $\numubar$ (antineutrino mode) is selectively generated, and $\sin\dcp$ is then measured by comparing the difference of the $P(\numu \to \nue)$ and $P(\numubar \to \nuebar)$ appearance probabilities. Other questions include the sign of $\Dmtt$, referred to as normal  ($\Dmtt > 0$, NO) or inverted ordering ($\Dmtt < 0$, IO), and whether $\ttt$ is maximal ($=\pi/4$), or in the upper or lower octant. These measurements have furthermore important consequences in the context of baryon asymmetry of the universe,\cite{leptogenesis} neutrinoless double beta decay searches,\cite{zntb} and flavor symmetries.\cite{flavorsym}

\section{The experiment}

To generate the neutrino beam, a $30\GeV$ proton beam is smashed into a $90\cm$ graphite target, which produces by hadronic interactions about 6~charged $\pi$ and $K$ mesons per proton. These are then focused by means of an optical system with three magnetic horns into a $96\un{m}$ long decay volume, where they subsequently decay in flight to produce neutrinos. By choosing the polarity of the horns, either positive $\pip,\kp$ or negative $\pim,\km$ particles are selected, which mostly produce $\numu$ or $\numubar$, respectively. The far detector is intentionally placed $2.5\,\mathrm{degrees}$ off the neutrino beam axis to obtain a narrowband beam peaked at the $650\MeV$ oscillation maximum.

A complex of multiple near detectors measures the neutrinos before oscillation. INGRID is an iron-scintillator sandwich detector that monitors the beam intensity and direction on-axis. ND280 precisely measures the neutrino flux and interactions at the same $2.5\deg$ off-axis angle as SuperK. It is a composite detector with active scintillator and passive water targets, with tracking in time projection chambers, and placed inside a magnet for charge and momentum measurement. A recent addition is the WAGASCI+BabyMIND\,\cite{babymind} detector system placed at an intermediate $1.5\deg$ off-axis angle. While not used for the oscillation analysis presented here, first neutrino cross-section measurements have been published recently.\cite{wagasci-xsec}

The far detector, SuperK,\cite{SuperK} is a $50\un{kt}$ water Cherenkov detector. The detector walls are lined with 11,146~photo-multiplier tubes (PMTs) that detect the Cherenkov light emitted from charged particles created in the neutrino interactions on water. It has very good $e/\mu$ PID from sharpness of the Cherenkov rings (mis-ID rate about 0.1\%), an ideal feature for a neutrino oscillation experiment measuring the appearance rate of $\nue$ in a $\numu$ beam. In T2K the incoming neutrino direction is known, such that by assuming quasi-elastic scattering the neutrino energy can be reconstructed from the charged lepton direction and momentum. Not being magnetized, the $\nu/\nubar$ separation relies on the beam composition, which is mostly $\numu$ or $\numubar$ in each operation mode. The wrong-sign background is constrained by ND280. Recently, Gadolinium loading into SuperK has started,\cite{skgd} which will allow for some $\nu/\nubar$ separation by tagging recoil neutrons captured on Gadolinium, most interesting for the search of supernova relic neutrinos.

\section{Oscillation analysis flow}

\subsection{Dataset}
In early 2020, we published constraints on $\dcp$ that excluded a wide range of its values at $3\sigma$ confidence level\,\cite{Nature2020} for the first time. The results presented in this work (also shown at Neutrino 2020\,\cite{Neutrino2020}) include additional data collected between 2019 and early 2020, corresponding to a 33\% increase in neutrino mode data. Stable operation at $515\,\mathrm{kW}$ beam power was achieved in this most recent run, which is the highest so far. The total amount of analyzed far detector data corresponds to $3.6\times 10^{21}$ protons on target (POT), with about 6:5 ratio of $\nu$\,:\,$\nubar$-mode POT.

\subsection{Flux prediction} 
Various information on primary protons from beam monitors is combined with dedicated hadron production data to predict the neutrino flux and uncertainties. A significant reduction of the hadron production uncertainty has been achieved in this analysis thanks to a change of the tuning method. Instead of tuning every single interaction with thin target measurements, we now use measurements taken with a replica of the T2K target\,\cite{NA61replica} and apply a single tuning weight based on the exiting particle's kinematics. Further reduction of the uncertainties are expected for the next analysis.

\subsection{ND280 constraints}
ND280 measurements are used to constrain the unoscillated flux $\times$ cross-section.
The interaction model has been updated significantly in this analysis, including the initial state nuclear model for the dominant charged current quasi-elastic (CCQE) scattering process, the treatment of removal energy, a new uncertainty model on the energy dependence of CC multi-nucleon knockout (2p2h) process, final state interaction constraints from the near detector, and improvements to deep inelastic scattering at higher energies.

To constrain this model, a total of 18~near detector samples are used, separated by neutrino/antineutrino mode, $\pi$ multiplicity (0, 1, ${\ge}\,2$) for interaction mode separation, lepton charge for constraining wrong-sign background (in antineutrino mode only), and interaction target for constraining nuclear effects (C from active scintillator, or C+O from active scintillator with inactive water layers). The amount of ND280 data included in the analysis has been doubled to $1.15\times 10^{21}$ ($0.834\times 10^{21}$) POT in neutrino (antineutrino) mode. Sample selections and systematics treatment have also been improved.  Two methods are used to propagate the near detector constraint to the far detector. One is to fit just the near detector samples and propagate a covariance matrix of correlated flux and cross section parameters. The other is to perform a joint fit of the near and far detector samples and thus avoid the gaussian approximation. Both methods give consistent results.

Aside from the oscillation analysis, many cross-section measurements have been published, including measurements of pionless $\numu$ charged-current interactions on C~vs.~O targets,\cite{cc0piCplusO} $\nu$~vs.~$\nubar$ differences,\cite{cc0pinunubar} $\nue/\nuebar$ cross section\,\cite{ccnuenuebar} (the first measurement of \,$\nuebar$ in 43~years), as well as single-pion interactions including proton tracks\,\cite{cc1pi1p} for understanding nuclear effects.

\begin{figure}
\begin{minipage}{0.49\linewidth}
\vspace{0.2em}
\centerline{\includegraphics[width=0.89\linewidth]{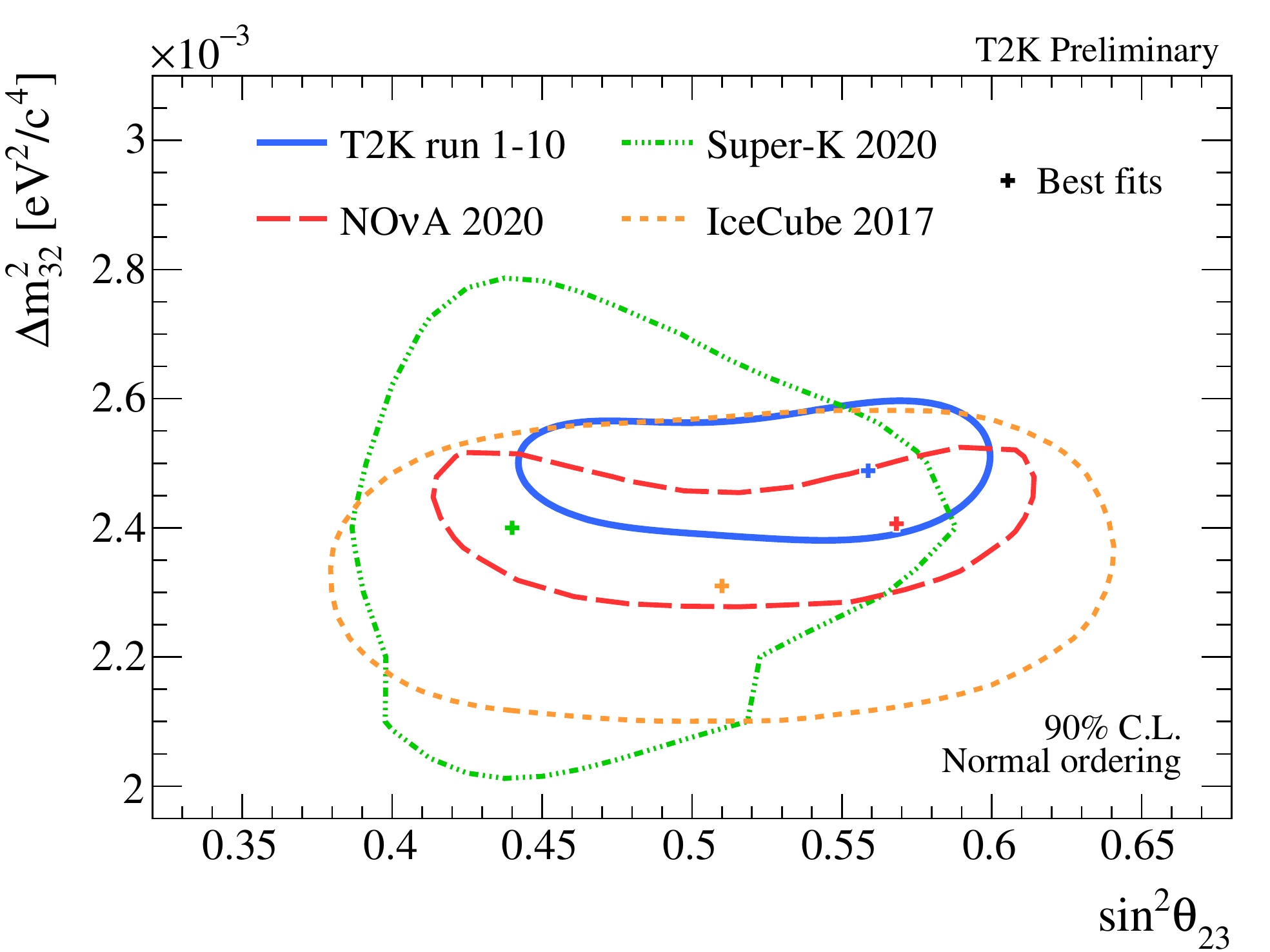}}
\end{minipage}
\hfill
\begin{minipage}{0.50\linewidth}
\centerline{\includegraphics[width=1\linewidth]{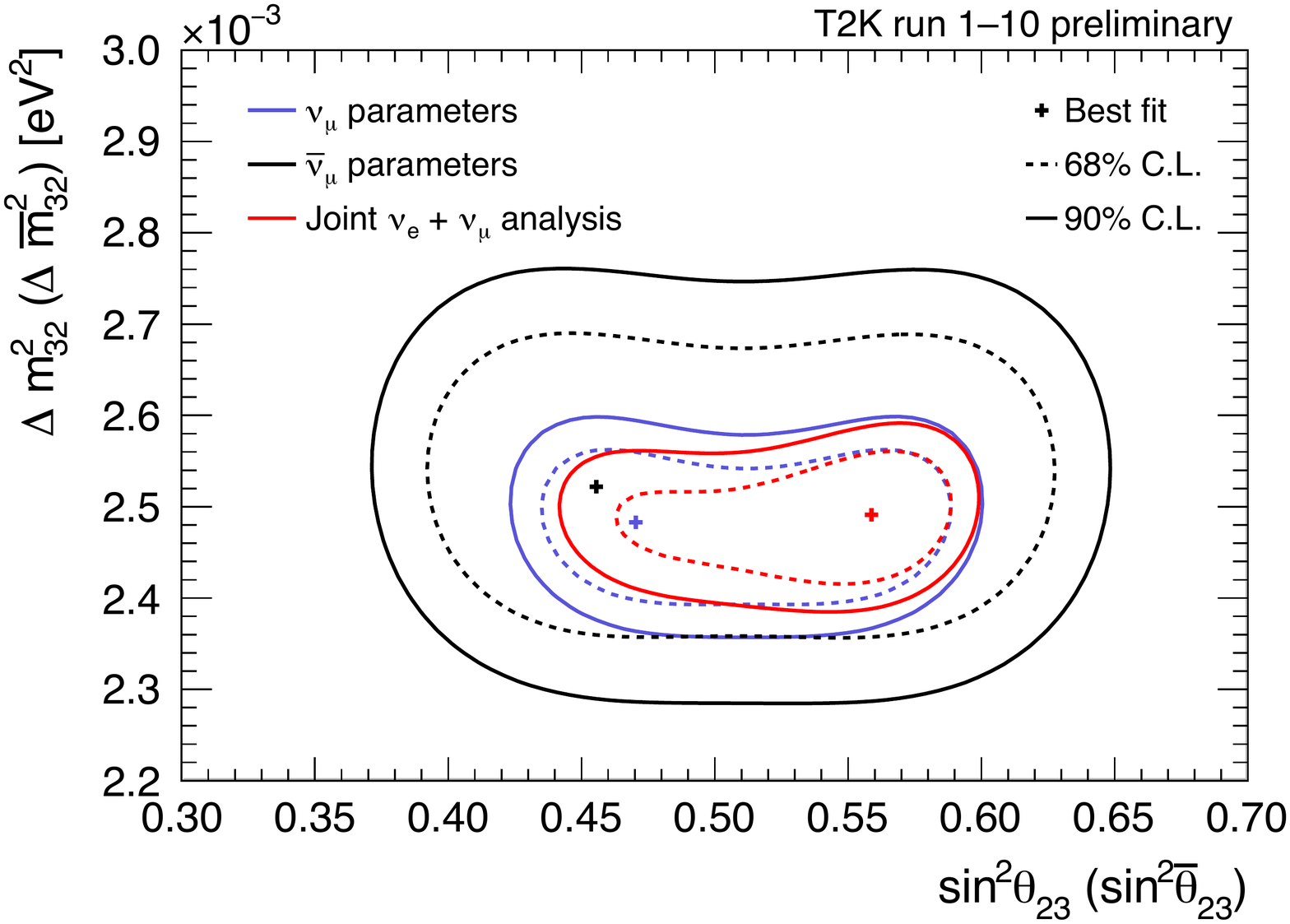}}
\end{minipage}
\caption[]{\emph{Left:} Constraints on atmospheric oscillation parameters and comparison with NO$\nu$A,\cite{nova2020} SuperK,\cite{superk2020} and IceCube.\cite{icecube2017} \emph{Right:} Constraints from an alternative fit of $\mu$-like samples where the atmospheric parameters for neutrinos $(\sintt,\Dmtt)$ and antineutrinos $(\sinttbar,\Dmttbar)$ are decoupled from each other. The joint $\nue+\numu$ analysis contours from the left figure are also shown for comparison.}
\label{fig:results:atm}
\end{figure}

\subsection{Far detector samples}
We use five single-ring samples separated by the beam mode and whether one sees an $e$-like or $\mu$-like ring. For neutrino mode, we also have a sample with single $e$-like ring and charged pion below Cherenkov threshold (detected as delayed event from decay electron) for increasing the $e$-like statistics and as control for CC1$\pi$ interactions.
% In this analysis, we extend the analysis binning of $\mu$-like samples from a 1D binning by reconstructed neutrino energy $E_\mathrm{rec}$ to a 2D binning $(E_\mathrm{rec},\theta_\mathrm{rec})$, where the reconstructed scattering angle $\theta_\mathrm{rec}$ allows improved separation from neutral current interaction backgrounds.
For this analysis, we extend the analysis binning of $\mu$-like samples by the muon scattering angle, which improves the separation of neutral current interaction backgrounds.
All data have been reprocessed for improved PMT gain correction.
%The data is reprocessed for improved PMT gain correction.
In the upcoming analysis we will further add multi-ring samples, which will provide up to 40\% increase of the $\mu$-like events (mostly at energies higher than oscillation maximum).

For $\mu$-like events, the disappearance rate at oscillation maximum is sensitive to $\sinttt$, whereas the energy at the dip is sensitive to $\Dmtt$. For $e$-like samples, a change of event rate correlated in $\nu/\nubar$-mode is sensitive to $\tot$ and $\ttt$-octant, whereas a change anti-correlated in $\nu/\nubar$-mode is sensitive to $\sin\dcp$ and mass ordering. T2K's $\tot$-constraint from $\numu\to\nue$ appearance, which is $\sinot = 2.63_{-0.42}^{+0.52}\times 10^{-2}$ (NO) in this analysis, is consistent with the much stronger constraint from reactor experiments measuring $\nuebar \to \nuebar$ disappearance. All following results therefore report T2K+reactor constraints on the other oscillation parameters $(\dcp,\ttt,\Dmtt)$, which improves the sensitivity especially for $\dcp$ and $\ttt$-octant. This reactor-constraint on $\tot$ has been updated since the last analysis to $\sinot = (2.18 \pm 0.07) \times 10^{-2}$ from Ref.~\cite{PDG}.

\section{Results}

Fig.~\ref{fig:results:atm} (left) shows our latest constraints on the atmospheric mixing parameters $(\ttt,\Dmtt)$. Improving upon our previous ones with increased statistics and systematics model, these are the tightest constraints on these parameters to date.
We observe a slight preference of non-maximal $\sintt \ne 0.5$ in upper octant with Bayes factor $P_\mathrm{upper}/P_\mathrm{lower} = 3.4$.
A fit with separate atmospheric mixing parameters for $\numu/\numubar$ (Fig.~\ref{fig:results:atm} right) shows consistent results as expected from CPT and standard neutrino interactions.

\begin{figure}
\begin{minipage}{0.49\linewidth}
\centerline{\includegraphics[width=0.9\linewidth]{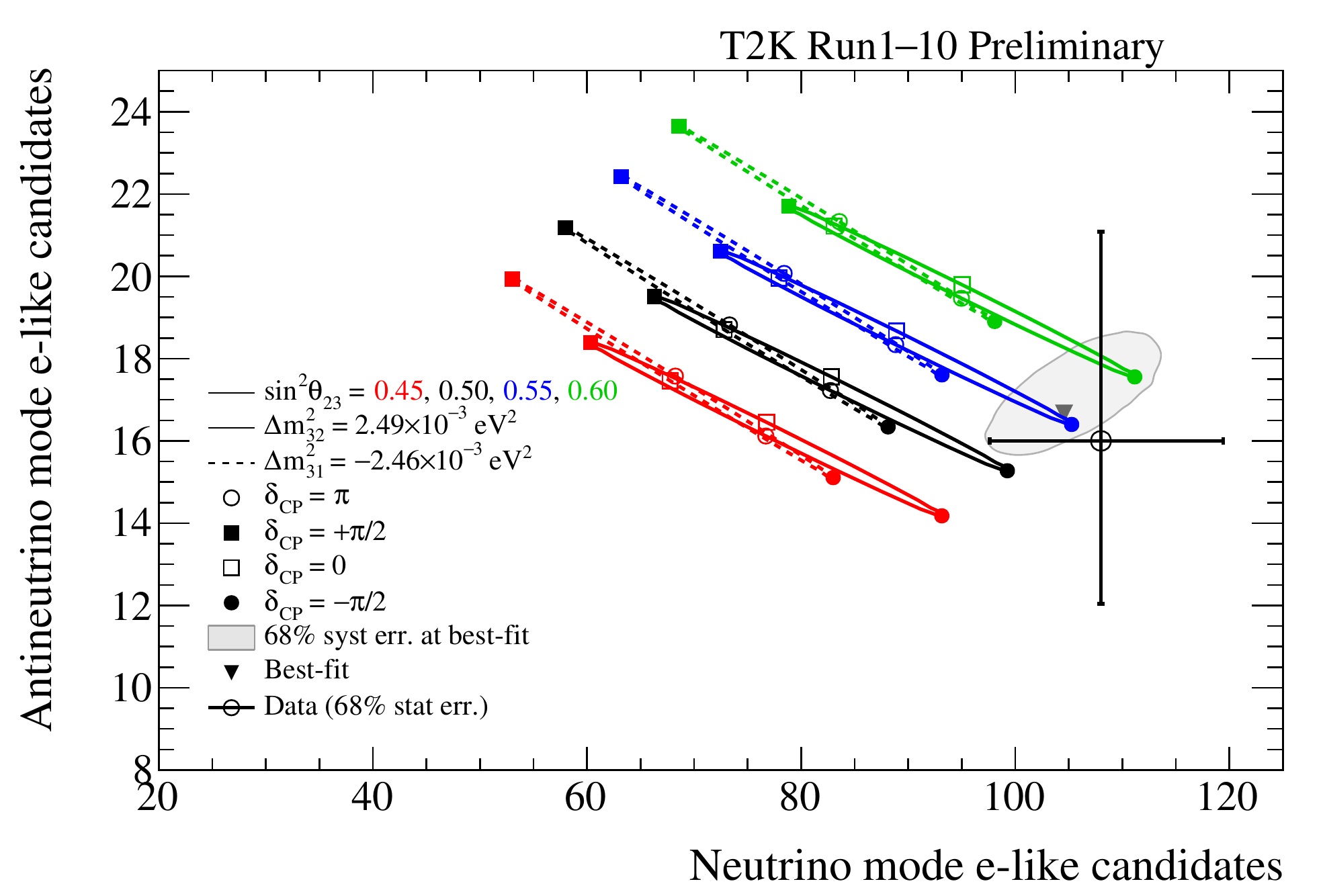}}
\end{minipage}
\hfill
\begin{minipage}{0.50\linewidth}
\centerline{\includegraphics[width=0.9\linewidth]{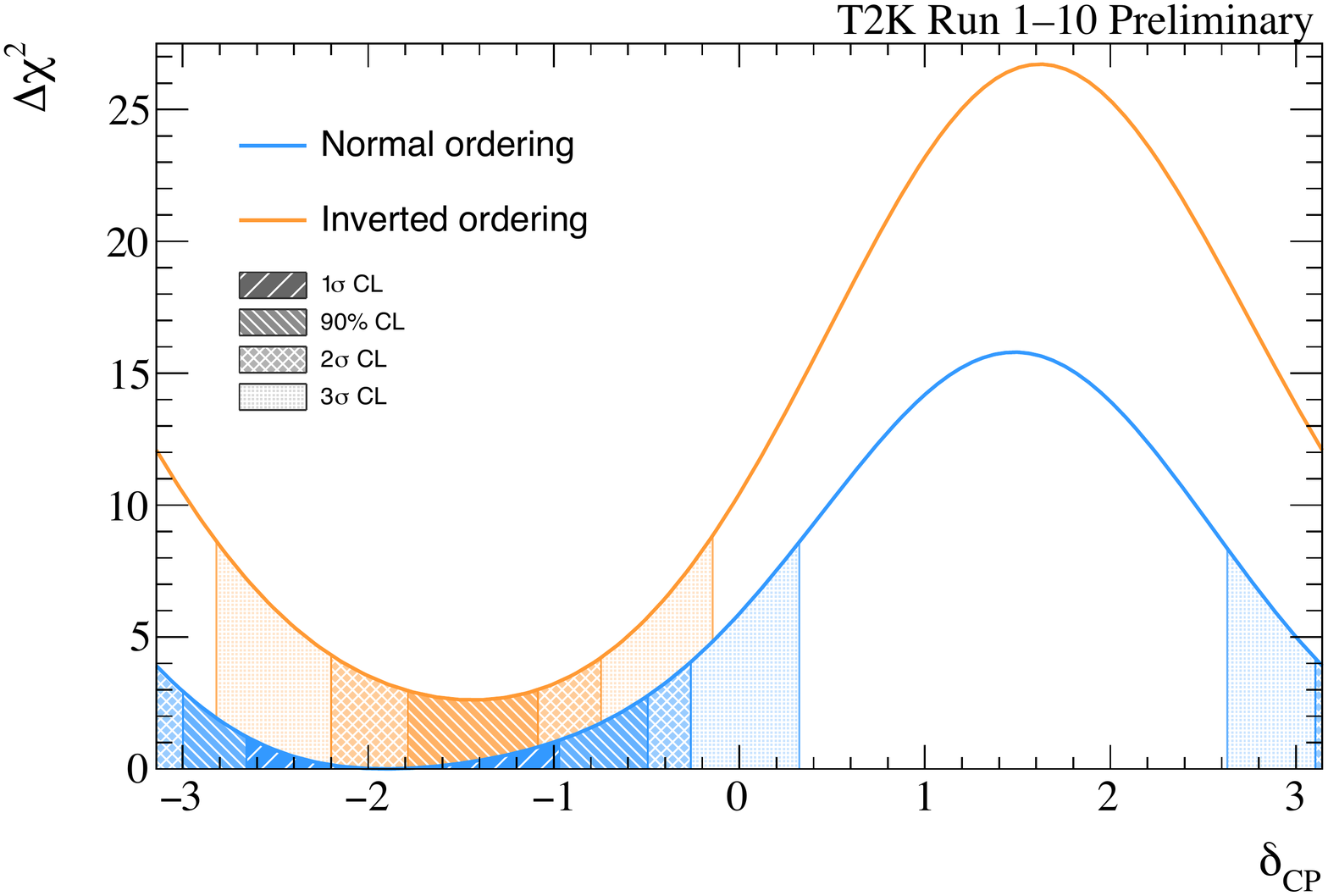}}
\end{minipage}
\caption[]{\emph{Left:} The ellipses show expected event rates for each value of $\dcp$, with data-point close to maximally CP-violating value of $-\pi/2$. The solid (dashed) ellipses show the prediction for normal (inverted) ordering, with our data showing weak preference for normal ordering. The colors show the prediction for various values of $\sintt$, with data showing a weak preference for upper octant. \emph{Right:} $\Dchi$ distribution for $\dcp$ in each mass ordering. The colored regions show the confidence intervals constructed with Feldman-Cousins method.\cite{feldmancousins}}
\label{fig:results:app}
\end{figure}

For $\nue$ vs. $\nuebar$ appearance the constraint is still rate-dominated, such that a bi-event plot is illustrative to understand the origin of various constraints (Fig.~\ref{fig:results:app} left). Upon statistical analysis, a large region of the $\dcp$ parameter space is excluded at $3\sigma$ (Fig.~\ref{fig:results:app} right), and CP-conserving values $\dcp=0,\pi$ are excluded at 90\% confidence level.
% To check the effects of possible biases caused by cross section model choices, we repeat the analysis with various theory or data-driven simulated data sets. The difference of the obtained $\Dchi$ distribution to the nominal model is superimposed onto the data-$\Dchi$, yielding a shift of the left (right) 90\% confidence interval edges on $\dcp$ by at most $0.073$ ($0.080$).
To check the effects of possible biases caused by cross section model choices, we repeat the analysis with various theory or data-driven simulated data sets. These cause a shift of the left (right) 90\% confidence interval edges on $\dcp$ by at most $0.073 \rad$ ($0.080 \rad$).
A weak preference of normal ordering is also observed with Bayes factor $P_\mathrm{NH}/P_\mathrm{IH} = 4.2$. Comparing with toy experiments generated at maximally CP-violating value of $\dcp = -\pi/2$, our data-constraints are compatible with sensitivity at around $1\sigma$ level.
In the previous analysis, the agreement used to be slightly below $2\sigma$ range, and was yielding by statistical fluctuation a stronger constraint than sensitivity. The new data-constraint is now \emph{weaker} due to the improvements to the analysis, including the updated interaction model and reactor-constraint on $\tot$, reprocessing of the far detector data, and with largest contribution the additional data. The sensitivity however increased, resulting in improved compatibility. 

% Since this additional data also caused an increase of sensitivity, the agreement between data-constraint and sensitivity has improved.

%Joint fits with other long-baseline neutrino oscillation experiments using different baselines, energies and detector technologies yield increased sensitivity in $\dcp$, mass ordering and $\ttt$-octant beyond increase in statistics due to resolution of oscillation parameter degeneracies and differences in systematics. To understand potentially non-trivial systematic correlations between these experiments, NO$\nu$A+T2K and SK+T2K joint fits with analyzers from each collaboration are ongoing.

\section{Conclusion and future outlook}

We presented latest neutrino oscillation results from T2K using $3.6\times 10^{21}$ protons on target with many improvements at each level of the analysis. CP conserving values of $\dcp$ are excluded at 90\% and a large $\dcp$ range is excluded at $3\sigma$ confidence level. A weak preference for normal ordering and upper octant is seen. Additional samples are being prepared both at near and far detectors for more statistics and systematics control. Joint fits between NO$\nu$A+T2K and SK+T2K collaborations are also ongoing, with the aim to obtain improved oscillation parameter constraints due to resolved degeneracies, and to understand potentially non-trivial systematic correlations.
Many upgrades are being performed toward the future, including a large upgrade of the ND280 near detector system for improved cross section model constraints,\cite{nd280upgrade} and $2.6\times$ stronger beam for significant increases in statistics.\cite{beamupgrade}

\end{document}